\documentclass[11pt]{article}
\usepackage{Blois,epsfig}

\def\be{\begin{equation}}
\def\ee{\end{equation}}
\def\bea{\begin{eqnarray}}
\def\eea{\end{eqnarray}}

\begin{document}

\title{ Searching the Color Glass Condensate Through $p_T$ Suppression}

\author{ M. A. Betemps \footnote{marcos.betemps@ufrgs.br}, 
         M. B. Gay Ducati \footnote{beatriz.gay@ufrgs.br} }

\address{High Energy Physics Phenomenology Group, GFPAE\\
Instituto de F\'{\i}sica, Universidade Federal do Rio Grande do Sul \\
Caixa Postal 15051, CEP 91501-970, Porto Alegre, RS, Brazil.}

\maketitle

\abstracts{
In this work the rapidity and transverse momentum dependence of the
nuclear modification ratio is evaluated for the dilepton production at
RHIC and LHC, using the Color Glass Condensate (CGC) framework. The
ratio is compared for two distinct dilepton mass values, concerning
the transverse momentum distribution, and a suppression of the Cronin
peak is verified even for large mass. The nuclear modification ratio
suppression in the dilepton rapidity spectra, as obtained for hadrons
at RHIC, is verified for LHC energies at large transverse momentum,
although not present at RHIC energies. These results consolidate the
dilepton as a most suitable observable for QCD high density
approaches.}

\section{Introduction}

The RHIC data on high energy collisions regarding the transverse
momentum spectra of the hadrons, present some interesting results
\cite{Debbe:2004ci,Arsene:2004ux} .  The comparison between
multiplicity of hadrons per unit of space in $d-Au$ and $p-p$, and in
central and peripheral collisions has being the main object of
investigation. These comparisons are performed introducing the nuclear
modification ratio, for example
\begin{eqnarray}
R_{dAu}=\frac{\frac{dN^{dAu\rightarrow
hX}}{dydp_T^2}}{N_{coll}\frac{dN^{pp\rightarrow hX}}{dydp_T^2}}, 
\end{eqnarray}
where the normalization factor $N_{coll}$ is the number of binary
collisions, $y$ and $p_T$ are the rapidity and transverse momentum of
the hadron.  For the case of $d-Au$ collisions the ratio becomes
larger than 1 at mid rapidity $y\approx 0$ and saturates at large
$p_T$. This implies a peak at intermediate $p_T$ (2-5 GeV), which is
called Cronin effect or Cronin peak.  As one goes to large rapidities,
the suppression of the ratio $R_{dAu}$ (smaller than 1) was observed
and a flat behavior at large $p_T$ is found. Such effects are measured
concerning the transverse momentum hadron spectra. In order to avoid
systematic errors, the ratio of a given centrality class to the
peripheral one should take $R_{CP}$ instead of the ratio
$R_{dAu}$. The peripheral nuclear collision behaves as a $pp$
collision and the ratio is obtained through the same run process and
same particle species.  In both defined ratios, the conclusions are
the same, a significant reduction of the yield of charged hadrons
measured in $dAu$ collisions when compared with the $pp$ collisions
at forward rapidities is found \cite{Arsene:2004ux}. Such suppression
effect is consistent with the expected saturation of the gluon
distribution, as proposed by the Color Glass Condensate approach when
quantum evolution is considered \cite{CGCsatCron}, and by the quantum
evolution of the BK equation \cite{Albacete:2003iq}.

The main interest related to the Cronin effect is regarding the hadron
$p_T$ spectra, however, in a recent work we have demonstrated that the
dilepton production should also be considered as an observable to
study this effect. The dileptons analyzed in the context of the Color
Glass Condensate (
e.g. \cite{MABMBGDCGC1,Jalilian-Marian:2004er,Baier:2004tj,Jalilian-Marian:2005jf})
present the same features of the Cronin effect in such approach
\cite{Blaizot:2004wu} and do no present final states 
interactions, carrying not disturbed information about the Color Glass
Condensate. In this work, the ratio between $p-Au$ and $p-p$
differential cross section is evaluated, analyzing the transverse
momentum and the rapidity distributions of the dileptons. Comparing
with the previous results
\cite{MABMBGDCGC1}, where the dilepton mass $M=3$ GeV was
employed, we calculate for larger values of mass and present the
results as a function of rapidity and transverse momentum $p_T$ of
the dileptons. We discuss the results and the kinematical limits
of the rapidity for this observable.

\section{Dilepton production in the CGC approach}

The dilepton production at high energies is dominated by the
bremsstrahlung of a virtual photon by a quark from a hadron
interacting with a dense background gluonic field of the nucleus,
and afterwards decaying into a lepton pair \cite{Gelis:2002ki}.

The Color Glass Condensate is a QCD classical effective theory to deal
with the high dense partonic system \cite{CGC}.  In this theory the
small $x$ gluons are described by a color source density $\rho_a$ and
radiated from fast moving color sources with internal dynamics frozen
by Lorentz time dilatation, thus forming a color glass. The
observables are obtained by means of an average over all
configurations of the color sources, performed through a weight
functional $W_{\Lambda^+}[\rho]$, which depends upon the dynamics of
the fast modes, and upon the intermediate scale $\Lambda^+$, which
defines fast ($p^+>\Lambda^+$) and soft ($p^+<\Lambda^+$) modes. The
modifications to the effective classical theory are governed by a
functional, nonlinear, evolution equation JIMWLK
\cite{RGE1,RGE2} for the statistical weight functional
$W_{\Lambda^+}[\rho]$ associated with the random variable $\rho_a(x)$.

The hadronic cross section of the dilepton production is obtained employing the
collinear factorization and considering the forward rapidity region,
and read as \cite{MABMBGDCGC1,dileptonGelisJJI},
\begin{eqnarray}
\frac{d\sigma^{pA\rightarrow
    ql^+l^-X}}{dp_T^2\,dM\,dx_F}=\frac{4\pi^2}{M} R^2_A
\frac{\alpha_{em}^2}{3\pi}\frac{1}{x_1+x_2}
\times \int \frac{dl_T}{(2\pi)^3} l_T
\,    {\cal W}(p_T,l_T,x_1)\,C(l_T,x_2,A),
\label{eqcsh}
\end{eqnarray}
with $x_F$ being the longitudinal momentum fraction given by
$x_F=x_1-x_2$ (related to the rapidity $y$), $M$ is the lepton pair mass,
$x_1$ and $x_2$ are the momentum fractions carried by the quark from
the proton and by the gluonic field from the nucleus, respectively,
defined in the formal way. The squared center of mass energy is
denoted by $s$ and $l_T=q_T+p_T$ is the total transverse momentum
transfer between the nucleus and the quark. The expression
(\ref{eqcsh}) is restricted to the forward region only, which means
positive rapidities $y$ (or $x_F$).  The function ${\cal
W}(p_T,l_T,x_1)$ can be written as \cite{dileptonGelisJJI},
\begin{eqnarray}\nonumber
{\cal W}(p_T,l_T,x_1)&=&\int_{x_1}^{1}dz\,z  F_2 (x_1/z,M^2)\\\nonumber
&&\times \left\{
 \frac{(1+(1-z)^2)z^2 l_T^2}{[ p_T^2+M^2(1-z)][( p_T-z
    l_T)^2+M^2(1-z)]}\right. \\
&\!\!-&\!\!\left.z(1-z)M^2\left[\frac{1}{[ p_T^2+M^2(1-z)]}
-\frac{1}{[(
    p_T-z l_T)^2+M^2(1-z)]}\right]^2 \right\}.
\label{cs2}
\end{eqnarray}
Here $R_A$ is
the nuclear radius,  $z\equiv p^-/k^-$ (light-cone variables) is the
energy fraction of the proton carried by the virtual photon.  The
function $C(l_T)$ is the field correlator function and defined by
\cite{Gelisqq1},
\begin{eqnarray}
C(l_T)\equiv \int d^2 x_{\perp} e^{il_T\cdot
  x_{\perp}}\langle U(0)U^{\dagger}(x_{\perp})\rangle _{\rho},
\label{defCk}
\end{eqnarray}
with the averaged factor representing the average over all
configurations of the color fields sources in the nucleus,
$U(x_{\perp})$ is a matrix in the $SU(N)$ fundamental representation
which represents the interactions of the quark with the classical
color field of the nucleus.  All the information about the nature of
the medium crossed by the quark is included in the function
$C(l_T)$. In particular, it determines the dependence on the
saturation scale $Q_s$ (and on energy), implying that all saturation
effects are encoded in this function. $F_2(x,Q^2)$ is the partonic
structure function, which takes into account the quark distribution of
the proton projectile.  In our calculations the CTEQ6L parametrization
\cite{Cteq6} was used for the structure function, and the lepton pair
mass gives the scale for the projectile quark distribution.

The energy dependence introduced in the Eq. (\ref{eqcsh}) in the
correlator function $C(l_T,x,A)$ is performed by means of the
saturation scale $Q_{s,A} (x)$ and provides the investigation of the
effect of the quantum evolution in the dilepton production. We are
based in the GBW parametrization \cite{GBW} to obtain the $x$
dependence of the saturation scale ($Q_s^2=(x_0/x)^{\lambda}$), and
the parameters have been taken from the dipole cross section extracted
from the fit procedure by CGCfit \cite{IIMunier} parametrization.  The
nuclear radius, which appears in the Eq.  (\ref{eqcsh}), is taken from
the Woods-Saxon parametrization, which has the form, $R_A=1.2 A^{1/3}$
$ fm$, while the proton radius (for $pp$ calculations) is taken from
the fits.

In this work we have evaluated the cross sections
using the function $C(l_T,x,A)$ based in the McLerran-Venugopalan (MV)
theory, introducing an $x$ dependence through the nuclear saturation
scale, which is parametrized in the form $Q_s ^2(x,A)=A^{1/3}Q_s^2$.

The correlator function employed here is a non-local Gaussian
distribution of color sources, predicted in Ref. \cite{Iancu:2001ad}
as a mean-field asymptotic solution for the JIMWLK equation written in
the following form \cite{Blaizot:2004wu,Gaussian},
\begin{eqnarray}
C(l_T,x,A) \equiv  \int d^2 x_{\perp} e^{il_T\cdot
  x_{\perp}}e^{-\chi(x,x_{\perp},A)},
\end{eqnarray}
with
\begin{eqnarray}\nonumber
\chi(x,x_{\perp},A)\equiv \frac{2}{\gamma c} \int
  \frac{dp}{p}(1-J_0(x_{\perp}p))
\times \ln\left(1+\left(\frac{Q_{2}^2(x,A)}{p^2}
\right)^{\gamma}\right),
\end{eqnarray}
where, $\gamma$ is the anomalous dimension ($\gamma\approx$ 0.64 for
BFKL) and $c\approx $ 4.84 \cite{Blaizot:2004wu,Gaussian}.

One has specified the cross section to evaluate the dilepton
differential cross section, and in the next section, the nuclear
modification ratio for the dilepton production is investigated and
related to the measured Cronin effect.

\section{Cronin effect and the dilepton production}

As discussed in the introduction of this work, the Cronin effect is
related to the measurement of the hadron transverse momentum
spectra. Here, the appearance of the same effects in the dilepton
$p_T$ and rapidity spectra are investigated for a lepton pair
mass $M=6$ GeV. In a previous work \cite{MABMBGDCGC1}, the
investigation was performed for the $p_T$ distribution of dileptons
with mass $M=3$ GeV.  Now, one evaluates the $p_T$ and rapidity
spectra for the dilepton at RHIC and LHC energies, $\sqrt{s}=200$
GeV and $\sqrt{s}=8800$ GeV, respectively.

We have defined the nuclear modification ratio for the dilepton
production in the following form,
\begin{eqnarray}
R_{pA}=
\frac{\frac{d\sigma(pA)}{\pi R_A^2dMdx_Fdp_T^2}}{A^{1/3}
\frac{d\sigma(pp)}{\pi R_p^2dMdx_Fdp_T^2}}.
\end{eqnarray}

\begin{figure}[ht]
\begin{center}
\scalebox{0.4}{\includegraphics{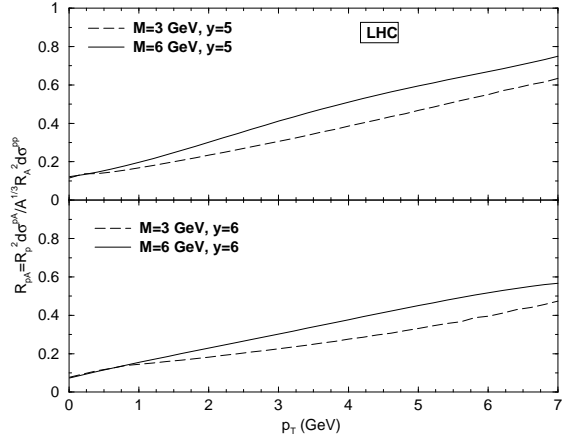}}
\caption{Ratio $R_{pA}$ for LHC energies, for $y=5$ and $y=6$,
comparing results for $M=3$ GeV and $M=6$ GeV.}
\label{ratioM6_3}
\end{center}
\end{figure}

The comparison between the results for the ratio $R_{pA}$ considering
two distincts lepton pair masses can be verified in the
Fig. \ref{ratioM6_3} for LHC energies, where the expected result is
found; the effect of the suppression of the ratio is reduced if the
dilepton mass is increased at a fixed rapidity.

\begin{figure}[ht]
\begin{center}
\scalebox{0.68}{\includegraphics{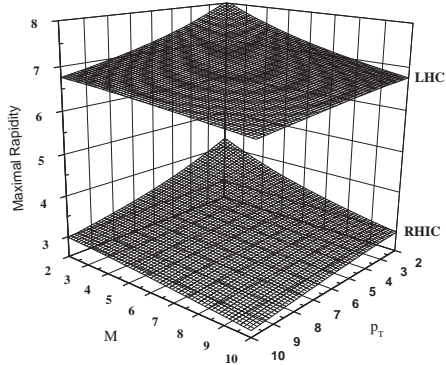}}
\caption{Maximal value of dilepton rapidity as a function of the mass
$M$ and transverse momentum $p_T$ for RHIC and LHC energies.}
\label{rapiditylimitRHICLHC}
\end{center}
\end{figure}

The performed analysis is only for forward rapidities. However, the
maximum value of rapidity depends on the value of the mass and the
transverse momentum. In the Fig. \ref{rapiditylimitRHICLHC} one
presents the limit values for the rapidity as a function of the
dilepton transverse momentum and mass for RHIC and LHC energies. The
region of large mass and large $p_T$ implies small values for the
rapidity limit values. In the RHIC kinematical regime, the mass region
between 2 and 10 GeV implies forward maximum values of rapidity from
2.97 to 2.65 respectively at $p_T\approx 10$ GeV and range between
4.26 to 2.97 respectively at $p_T \approx 2$ GeV for an energy
$\sqrt{s}=200$ GeV.

For LHC energies the same behavior is verified, however, the range of
the maximum rapidity is dramatically modified. For the mass region
between 2 and 10 GeV, the maximum rapidity range goes from 6.76 to
6.43 respectively at $p_T\approx 10$ GeV and goes from 8.04 to 6.76
respectively at $p_T\approx 2$ GeV for an energy $\sqrt{s}=8800$ GeV.
The nuclear modification ration, is investigated up to the maximal
value of the forward rapidity for mass $M=6$ GeV at RHIC and LHC
energies.

\begin{figure}[ht]
\begin{center}
\scalebox{0.8}{\includegraphics{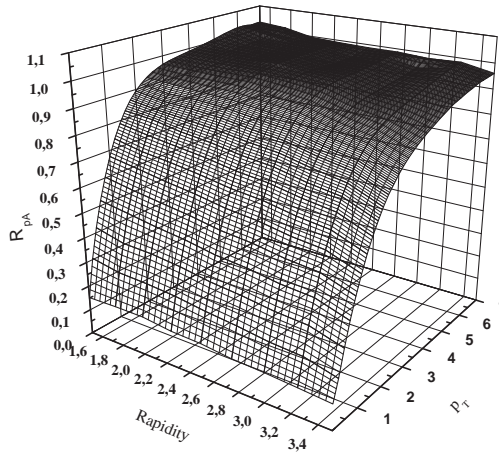}} 
\caption{Ratio $R_{pA}$ as a function of rapidity and $p_T$ for 
dileptons at RHIC energies.} \label{ratio3DRHIC}
\end{center}
\end{figure}

In the Fig. \ref{ratio3DRHIC} the nuclear modification ratio for RHIC
energies is shown for dilepton mass $M=6$ GeV.  The weak dependence of
the ratio $R_{pA}$ with the rapidity range for the RHIC energies is
verified, since independent of the $p_T$ value, the ratio does not
vary significantly with rapidity. This occurs due to the fact that one
evaluates the ratio $R_{pA}$ only at forward rapidities. The
suppression of the ratio with the increase of the rapidity is verified
concerning the hadron spectra from mid rapidity ($y=0$) to forward
rapidity ($y=3.2$) \cite{Arsene:2004ux}. In the case of the dileptons,
the same suppression is verified, however as we are restricted to the
forward rapidities, such suppression is small in the rapidity range
investigated here. For completeness, in such range of rapidity
(1.5$<y<$3) and for $M=6$ GeV, the nuclear dense system is being
proved with the momentum fraction from $x=10^{-2}$ down to $x=10^{-3}$
at $p_T\approx 10$ GeV, being not too small $x$ region.  The
suppresion of the ratio (absence of a Cronin type peak) in the $p_T$
distribution is verified, although the suppression is pratically the
same as one goes up in rapidity.

\begin{figure}[ht]
\begin{center}
\scalebox{0.8}{\includegraphics{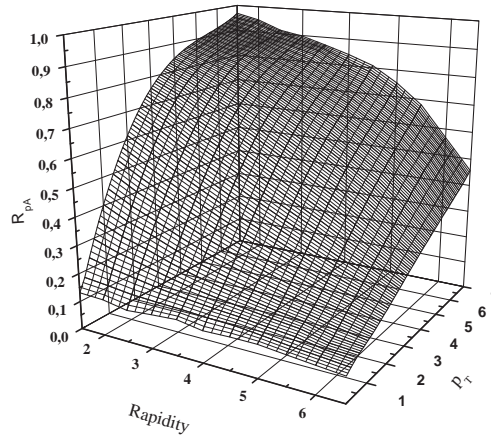}}
\caption{Ratio $R_{pA}$ as a function of rapidity and $p_T$ for dileptons
at LHC energies.}
\label{ratio3DLHC}
\end{center}
\end{figure}

The comparison with the previous ratio results evaluated for mass
$M=3$ GeV \cite{MABMBGDCGC1}  was done in the Fig. \ref{ratioM6_3} for
LHC energies.  In the Fig. \ref{ratio3DLHC} the nuclear modificaton
ratio for LHC energies is shown for dilepton mass $M=6$ GeV. Due to
the large range of forward rapidities at LHC energies, one verifies
the large suppression of the nuclear modification ratio with the
increase of the rapidity. This suppression is intensified at large
$p_T$.  The suppression of the same ratio with the transverse momentum
is also verified and is intensified at large rapidities.

At LHC, the large range of rapidity provides the $x$ range in the
large $p_T$ region ($p_T\approx 10$ GeV) stays between $10^{-4}$ and
$10^{-6}$. In this kinematical region, there are large effects of
saturation predicted by the Color Glass Condensate: the large
suppression of the nuclear modification ratio comparing with the
expected Cronin peak shows the existence of the saturation effects, in
both, rapidity and transverse momentum distributions.

In this work the ratio $R_{pA}$ was investigated for dilepton mass
$M=6$ GeV and shown as a function of rapidity and $p_T$. One has
discussed the region of maximum value of rapidity as a function of the
dilepton mass and transverse momentum. The results presented here,
show that the dilepton should be considered as a good observable to
provide a tool to investigate the properties of the Cronin effect,
since it presents the effect of suppression of the ratio $R_{pA}$ in
both distributions, rapidity and transverse momentum, and does not
present final state interactions.  One expects that such saturation
effects are from the similar mechanisms observed in the hadrons
transverse momentum and rapidity spectra, and considers the dilepton
production should clarify the status of final and initial state
effects in the Cronin effect being a cleanest probe to the Color Glass
Condensate dynamics.

{\it Acknowledgments:} MBGD would like to thank the organizers of the
XVIIth Rencontre de Blois for the invitation. This work was
supported by the CNPq, Brazil.

\end{document}